\documentclass[conference]{IEEEtran}
\IEEEoverridecommandlockouts
\usepackage{cite}
\usepackage{amsmath,amssymb,amsfonts}
\usepackage{algorithmic}
\usepackage{graphicx}
\usepackage{textcomp}
\usepackage{xcolor}
\usepackage{subfigure}
\usepackage{comment}
\usepackage{adjustbox}
\usepackage{multirow,tabularx}
\usepackage{array,etoolbox}
\preto\tabular{\setcounter{magicrownumbers}{0}}
\newcounter{magicrownumbers}

\usepackage{xcolor,colortbl}
\definecolor{codegreen}{rgb}{0,0.6,0}
\definecolor{codegray}{rgb}{0.5,0.5,0.5}
\definecolor{codepurple}{rgb}{0.58,0,0.82}
\definecolor{backcolour}{rgb}{0.95,0.95,0.92}
\definecolor{LightBlue}{rgb}{0.83, 0.91, 1}
\definecolor{LightGreen}{rgb}{0.8, 1, 0.8}
\definecolor{LightPink}{rgb}{1, 0.8, 0.88}
\definecolor{LightYellow}{rgb}{1, 1, 0.6}
\definecolor{light-gray}{gray}{0.9}

\newcommand{\wname}{BTHOWeN}
\definecolor{revhl}{RGB}{200,0,20}
\newcommand{\revision}[1]{{\textcolor{revhl}{#1}}}
\renewcommand{\revision}[1]{{#1}} 

\def\BibTeX{{\rm B\kern-.05em{\sc i\kern-.025em b}\kern-.08em
    T\kern-.1667em\lower.7ex\hbox{E}\kern-.125emX}}
\begin{document}


\title{Weightless Neural Networks for Efficient Edge Inference}


\author{
\IEEEauthorblockN{Zachary Susskind\IEEEauthorrefmark{1}, 
Aman Arora\IEEEauthorrefmark{2}}
\IEEEauthorblockA{
\textit{Department of Electrical and Computer Engineering} \\
\textit{The University of Texas at Austin}
}
\IEEEauthorblockN{
Igor Dantas Dos Santos Miranda\IEEEauthorrefmark{3}}
\IEEEauthorblockA{
\textit{Department of Electrical and Computer Engineering} \\
\textit{Federal University of Recôncavo da Bahia}
}
\IEEEauthorblockN{
Luis Armando Quintanilla Villon\IEEEauthorrefmark{4},
Rafael Fontella Katopodis}
\IEEEauthorblockA{
\textit{Department of Electrical and Computer Engineering} \\
\textit{Federal University of Rio de Janeiro}
}
\IEEEauthorblockN{
Leandro Santiago de Araújo}
\IEEEauthorblockA{
\textit{Department of Electrical and Computer Engineering} \\
\textit{Universidade Federal Fluminense}
}
\IEEEauthorblockN{
Diego Leonel Cadette Dutra,
Priscila Machado Vieira Lima,
Felipe Maia Galvão França}
\IEEEauthorblockA{
\textit{Department of Electrical and Computer Engineering} \\
\textit{Federal University of Rio de Janeiro}
}
\IEEEauthorblockN{
Mauricio Breternitz Jr.}
\IEEEauthorblockA{
\textit{ISTAR } \\
\textit{ISCTE Instituto Universitario de Lisboa}
}
\IEEEauthorblockN{
Lizy K. John
}
\IEEEauthorblockA{
\textit{Department of Electrical and Computer Engineering} \\
\textit{The University of Texas at Austin}
}
\vspace{-2ex}
}


\maketitle

\begin{abstract}
Weightless Neural Networks (WNNs) are a class of machine learning model which use table lookups to perform inference. This is in contrast with Deep Neural Networks (DNNs), which use multiply-accumulate operations.
State-of-the-art WNN architectures have a fraction of the implementation cost of DNNs, but still lag behind them on accuracy for common image recognition tasks. Additionally, many existing WNN architectures suffer from \revision{high memory requirements}.
In this paper, we propose a novel WNN architecture, \wname, \revision{with key algorithmic and architectural improvements over prior work}, namely  \revision{counting} Bloom filters, \revision{hardware-friendly hashing}, 
and \revision{Gaussian-based nonlinear} thermometer encodings to improve model accuracy and reduce area and energy consumption. \wname~targets the large and growing edge computing sector by providing superior latency and energy efficiency to comparable quantized DNNs.
Compared to state-of-the-art WNNs across nine classification datasets, \wname~on average reduces error by more than than 40\% and model size by more than 50\%.
We \revision{then demonstrate the viability of the \wname~architecture by presenting an} FPGA-based \revision{accelerator}, and compare its latency and resource usage against similarly accurate \revision{quantized} DNN accelerators, including Multi-Layer Perceptron (MLP)  and convolutional models.
The proposed \wname~models consume almost 80\% less energy than the MLP models, with nearly 85\% reduction in latency. In our quest for efficient ML on the edge, WNNs are clearly deserving of additional attention.
\end{abstract}


\section{Introduction}
\label{sec:introduction}

In the last decade, Deep Neural Networks (DNNs) have driven revolutionary improvements in the accuracy of tasks such as image recognition, image classification, speech recognition, and natural language processing. In fact, it is widely acknowledged that modern DNNs can achieve superhuman accuracy on image recognition and classification tasks\cite{russa2015}. However, the implementation of these models is expensive in both memory and computation. Table \ref{tab:dnnsize} shows the number of weights and multiply-accumulate operations (MACs) needed for some widely-known networks. These networks have excellent accuracy, but performing inference  with them requires significant memory capacity and MAC computation, which in turn consumes a substantial amount of energy. This may be acceptable on large servers, but in the emerging domain of edge computing, models must be run on small, power-constrained devices. The amount of weight memory and the number of computations required by these DNNs make them impractical to implement in edge solutions. Consequently, DNNs for edge inference must typically trade off accuracy for reduced complexity through techniques such as pruning and low-precision quantization \cite{gholami2021survey}.

Weightless Neural Networks (WNNs) are an entirely distinct class of neural model, inspired by the decode processing of input signals received by the dendritic trees of biological neurons~\cite{wnn_intro_esann}.
WNNs are composed of artificial neurons with discrete (usually binary) inputs and outputs which do not use weights to determine their responses. Instead, WNN neurons, also known as \textit{RAM nodes}, use Lookup Tables (LUTs) to represent Boolean functions of their inputs as truth tables. Rather than performing arithmetic operations with their inputs, RAM nodes simply concatenate them to form an address, then perform a table lookup to determine their response. A RAM node with $n$ inputs can represent any of the $2^{2^n}$ possible logical functions of its inputs using $2^{n}$ bits of storage.

\begin{table}[htbp]
\centering
\caption{Weights and MACs for popular DNNs~\cite{vivienne17} \cite{mythic}}
\begin{tabular}{|c|c|c|c|c|c|} 
 \hline
 \rowcolor{LightBlue} \centering
 Metric & LeNet-5 & AlexNet & VGG-16 & Resnet-50 & OpenPose\\
 \hline
 \#Weights & 60k & 61M & 138M &  25.5M & 46M \\
 \hline
 \#MACs & 341k & 724M & 15.5G & 3.9G & 180G\\
 \hline
 Year & 1998 & 2012 & 2014 & 2015 & 2018\\
 \hline
\end{tabular}
\label{tab:dnnsize}
\end{table}

Foundational research in WNNs occurred from the 1950s through the 1970s. However, WiSARD (Wilkie, Stonham, and Aleksander’s Recognition Device)~\cite{wisard}, introduced in 1981 and sold commercially from 1984, was the first WNN to be broadly viable. WiSARD was a pattern recognition machine, specialized for image recognition tasks. Two factors led to its success. First, then-recent advancements in integrated circuit manufacturing allowed for the fabrication of complex devices with large RAMs. Additionally, WiSARD incorporated algorithmic improvements which greatly increased its memory efficiency over simpler WNNs, allowing for the implementation of more sophisticated models. As recent results have formally shown, the VC dimension\footnote{The 
Vapnik–Chervonenkis (VC) dimension measures the complexity of the knowledge represented by a set of functions that can be encoded by a binary classification algorithm\cite{vc_dimension}. While usually approximated by statistical methods, it is possible to establish the exact VC dimension for some learning methods, including WiSARD.} of WiSARD is very large \cite{wisard_vc}, meaning it has a large theoretical capacity to learn patterns. Many subsequent WNNs\cite{weightless_review}, including the model proposed in this paper, draw inspiration from WiSARD's basic architecture.

Training a WNN entails learning logical functions in its component RAM nodes. Both supervised \cite{10.5555/284803.284804} and unsupervised \cite{unsupervised_wnn} learning techniques have been explored for this purpose. Many training techniques for WNNs directly set values in the RAM nodes. The mutual independence between nodes when LUT entries are changed means that each input in the training set only needs to be presented to the network once. By contrast, most DNN training techniques involve iteratively adjusting weights, and many epochs of training may be needed before a model converges. By leveraging one-shot training techniques, WNNs can be trained up to four orders of magnitude faster than DNNs and other well-known computational intelligence models such as SVM~\cite{cluswisard}.

Algorithmic and hardware improvements, combined with widespread research efforts, drove rapid and substantial increases in DNN accuracies during the 2010s. The ability to rapidly train large networks on powerful GPUs and the availability of big data fueled an AI revolution which is still taking place. 
While DNNs drove this revolution, we believe that WNNs are now a concept worth revisiting due to increasing interest in low-power edge inference.
WNNs also have potential as tools to accompany DNNs. For instance, it has been demonstrated that WNNs can be used to dramatically speed up the convergence of DNNs during training \cite{hybrid_wnn}. There are also many applications where a small network is run first for approximate detection; then, if needed, a larger network is used for more precision\cite{kang2021accelerating}. The approximate networks by design do not need high accuracy; high speed and low energy usage are more important considerations. WNNs are perfect for these applications.
However, in order to realize the benefits of this class of neural network, work needs to be done to design optimized WNNs with high accuracy and low area and energy costs.

Microcontroller-based approaches to edge inference, such as tinyML, have attracted a great deal of interest recently due to their ability to use inexpensive off-the-shelf hardware\cite{arduino}. However, these approaches to machine learning are thousands of times slower than dedicated accelerators.


In this paper, we explore techniques to improve the accuracy and reduce the hardware requirements of WNNs. These techniques include \revision{ hardware-efficient counting Bloom filters, hardware implementation of} recent algorithmic improvements such as bleaching \cite{Grieco:2010:PPE:1751674.1751890}\cite{bleaching}\revision{, and a novel} nonlinear thermometer encoding. We combine these techniques to create a software model and hardware architecture for WNNs which we call \wname~(\textbf{B}leached \textbf{T}hermometer-encoded \textbf{H}ashed-input \textbf{O}ptimized \textbf{We}ightless Neural \textbf{N}etwork; pronounced as \textit{Beethoven}). We present FPGA implementations of inference accelerators for this architecture, discuss their associated tradeoffs, and compare them against prior work in WNNs and against DNNs with similar accuracy.

Our specific contributions in this paper are as follows:
\begin{enumerate}
    \item \wname, a weightless neural network architecture designed for edge inference, which incorporates \revision{novel,} hardware-efficient \revision{counting} Bloom filters, \revision{nonlinear} thermometer encoding, and bleaching.
    \item Comparison of \wname~with state-of-the-art WiSARD-based WNNs across nine datasets, with a mean 41\% reduction in error and 51\% reduction in model size.
    \item An FPGA implementation of the \wname~architecture, which we compare against MLP and CNN models of similar accuracy on the same nine datasets, finding a mean 79\% reduction in energy and 84\% reduction in latency versus MLP models. Compared to CNNs of similar accuracy, the energy reduction is over 98\% and latency reduction is over 99\%.
    \item A toolchain for generating \wname~models, including automated hyperparameter sweeping and bleaching value selection. A second toolchain for converting trained \wname~models to RTL for our accelerator architecture. These are available at: \textit{URL omitted for double-blinding}.

\end{enumerate}

The remainder of our paper is organized as followed: In Section \ref{sec:background}, we provide additional background on WNNs, WiSARD, and \revision{prior} algorithmic improvements. In Section \ref{sec:proposal}, we present the \wname~ architecture in detail. In Section \ref{sec:methodology}, we discuss software and hardware implementation details. In Section \ref{sec:results}, we compare our model architecture against prior memory-efficient WNNs, and compare our accelerator architecture against a prior WNN accelerator and against MLPs and CNNs of comparable accuracy. Lastly, in Section \ref{sec:conclusion}, we discuss future work and conclude. 
\section{Background and Prior Work}
\label{sec:background}


\subsection{Weightless Neural Networks}
Weightless neural networks (WNNs) are a type of neural model which use table lookups for computation. WNNs are sometimes considered a type of Binary Neural Network (BNNs), but their method of operation differs significantly from other BNNs. Most BNNs are based around popcounts\revision{,} i.e. counting the number of 1s in some bit vector. For instance, the McCulloch-Pitts neuron\cite{McCulloch1943-MCCALC-5}, one of the oldest and simplest neural models, performs a popcount on its inputs and compares the result against a fixed threshold in order to determine its output. More modern approaches first take the XNOR of the input with a learned weight vector, allowing an input to be negated before the popcount occurs\cite{courbariaux2016binarized}.

The fundamental unit of computation in WNNs is the RAM node, an $n$-input, $2^{n}$-output lookup table with learned 1-bit entries. Conventionally, all entries in the RAM nodes are initialized to 0. During training, inputs are binarized or discretized using some encoding scheme and then presented to the RAM nodes. The input bits to a node are concatenated to form an address, and the corresponding entry in the node's LUT is set to 1.
Note that presenting the same input to the node again has no effect, since the corresponding bit position has already been set. Therefore, an advantage of this approach is that each training sample only needs to be presented once.

Lookup tables are able to implement any Boolean function of their inputs. Therefore, in theory, a WNN can be constructed with a single RAM node which takes all (encoded) input features as inputs. However, this approach has two major issues. First, the size of a RAM node grows exponentially with its number of inputs. Suppose we take a dataset such as MNIST\cite{lecun-mnisthandwrittendigit-2010} and apply a simple encoding strategy such that each of the original inputs is represented using 1 bit. Since images in the MNIST dataset are 28x28, our input vector has 784 bits, and therefore the RAM node requires $2^{784}$ bits of storage, about $1.7 * 10^{156}$ times the number of atoms in the visible universe. 
The second issue is that RAM nodes have no ability to generalize: if a single bit is flipped in an input pattern, the node can not recognize it as being similar to a pattern it has seen before.

A great deal of WNN literature revolves around finding solutions to these two issues. A discussion of many of these approaches can be found in \cite{wnn_intro_esann} \cite{weightless_review} . Unfortunately, many of these techniques require random behavior (e.g. replacing the entries in RAM nodes with Bernoulli random variables), which is challenging to implement in hardware\revision{.} The WiSARD model addresses both issues with the single-RAM-node model while avoiding the pitfalls of other solutions.

\revision{There are some structural similarities between WNNs and architectural predictors in microprocessors. For instance, using a concatenated input vector to index into a RAM node is conceptually similar to using a branch history register in a table-based branch predictor.}

\subsection{WiSARD}
\begin{figure}[hbtp]
\centerline{\includegraphics[width = 1.0\columnwidth]{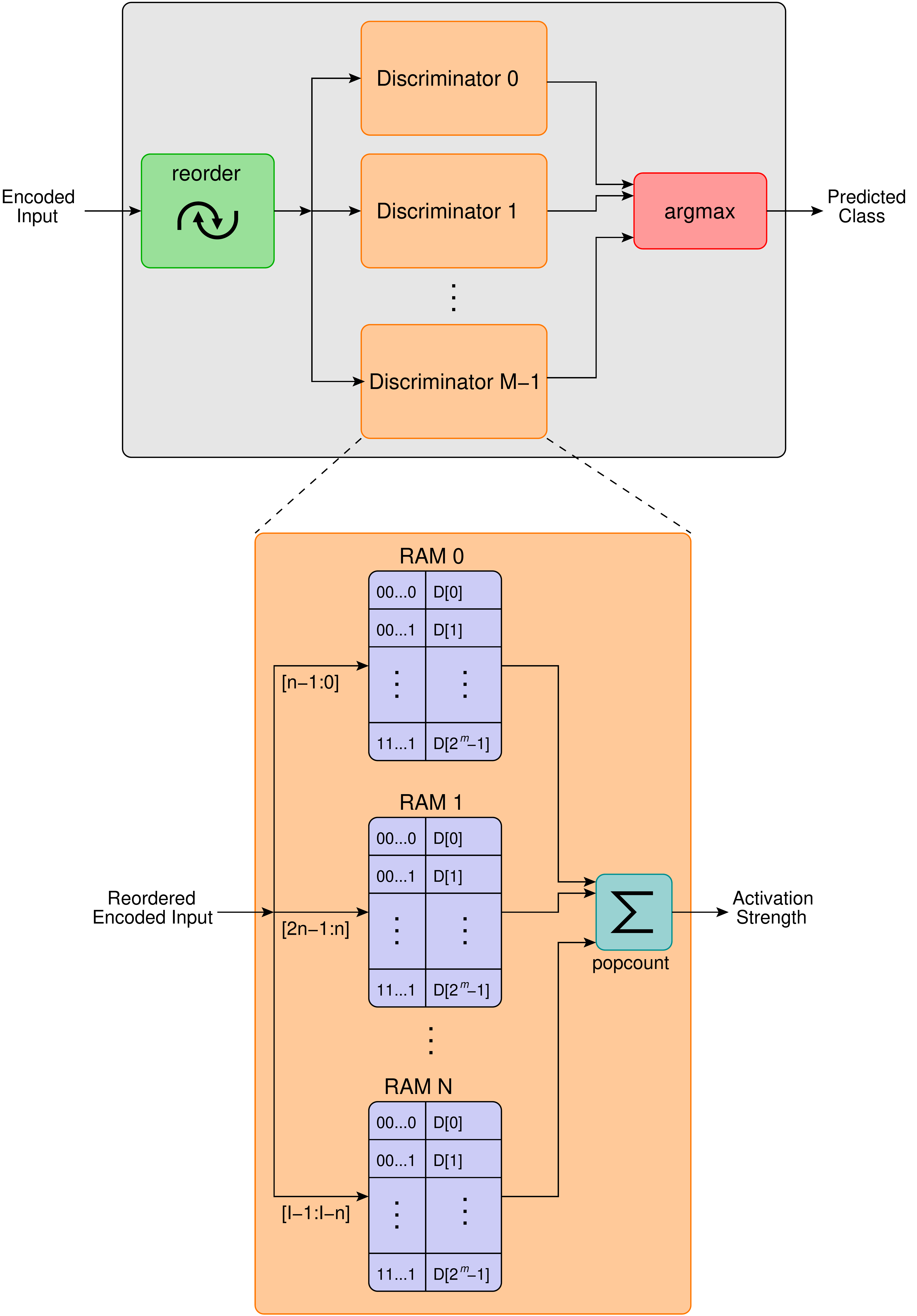}}
\caption{A depiction of the WiSARD WNN model with $I$ inputs, $M$ classes, and $n$ inputs per RAM node. $I/n$ RAM nodes are needed per discriminator, for a total of $M(I/n)$ nodes and $M(I/n)2^{n}$ bits of state.}
\label{fig:wisard}
\end{figure}

WiSARD\cite{wisard}, depicted in Figure \ref{fig:wisard}, is perhaps the most broadly successful weightless neural model. WiSARD is intended primarily for classification tasks, and has a submodel known as a \textit{discriminator} for each output class. Each discriminator is in turn composed of $n$-input RAM nodes; for an $I$-input model, there are $N \equiv I/n$ nodes per discriminator. Inputs are assigned to these RAM nodes using a pseudo-random mapping; typically, as in Figure \ref{fig:wisard}, the same mapping is shared between all discriminators.

\begin{figure}[hbtp]
\centerline{\includegraphics[width = 1.0\columnwidth]{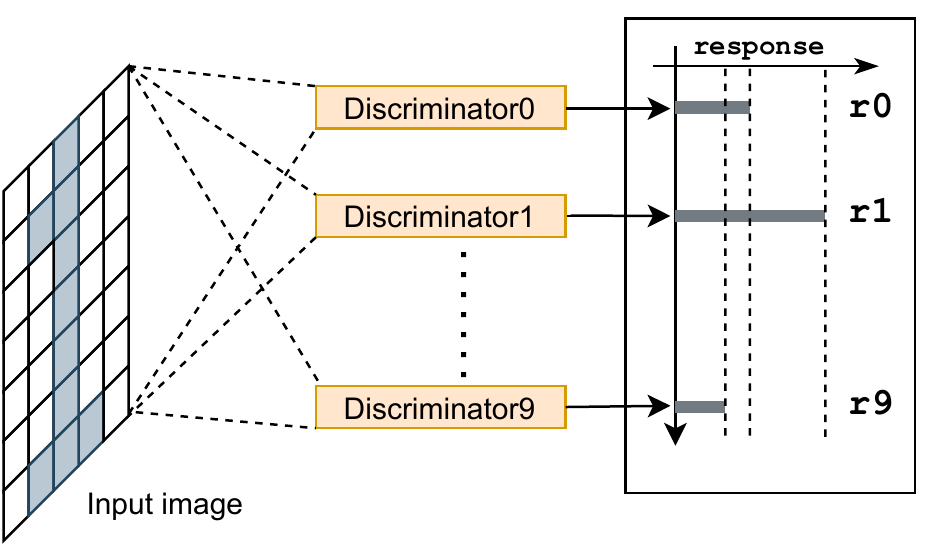}}
\caption{A WiSARD model recognizing digits. In this example, the input digit is 1, and the corresponding discriminator produces the strongest response.}
\label{fig:wisard_basics}
\end{figure}

During training, inputs are presented only to the discriminator corresponding to the correct output class, and its component RAM nodes are updated. During inference, inputs are presented to \textit{all} discriminators. Each discriminator then forms a bit vector from the outputs of its component RAM nodes and performs a popcount on this vector to produce a response value. The index of the discriminator with the highest response is taken to be the predicted class. Figure \ref{fig:wisard_basics} shows a simplified view of a WiSARD model performing inference. The response from Discriminator 1 is the highest since the input image contains the digit ``1".

If an input seen during inference is identical to one seen during training, then all RAM nodes of the corresponding discriminator will yield a 1, resulting in the maximum possible response. On the other hand, if a pattern is similar but not identical, then some subset of the RAM nodes may produce a 0, but many will still yield a 1. As long as the response of the correct discriminator is still stronger than the responses of all other discriminators, the network will output a correct prediction. In practice, WiSARD has a far greater ability to generalize than simpler WNN models.

WiSARD's performance is directly related to the choice of $n$. Small values of $n$ give the model a great deal of ability to generalize, while larger values produce more specialized behavior. On the other hand, larger values of $n$ increase the complexity of the Boolean functions that the model can represent \cite{wisard}.


\subsection{Bloom Filters}
Although the WiSARD model avoids the state explosion problem inherent in large, simple WNNs, practical considerations still limit the sizes of the individual RAM nodes. Increasing the number of inputs to each RAM node will, up to a point, improve the accuracy of the model; however, the model size will also increase exponentially.
Fortunately, the contents of these large RAM nodes are highly sparse, as few distinct patterns are seen during in training relative to the large number of entries available. Prior work has shown that using hashing to map large input sets to smaller RAMs can greatly decrease model size at a minimal impact to accuracy\cite{Arajo2019MemoryEW}.

A Bloom Filter\cite{BloomFilter} is a hash-based data structure for approximate set membership. When presented with an input, a Bloom filter can return one of two responses: 0, indicating that the input is definitely not a member of the set, or 1, indicating that the element is \textit{possibly} a member of the set. False negatives do not occur, but false positives can occur with a probability that increases with the number of elements in the set and decreases with the size of the underlying data structure\cite{bloom_false_positive}.
Bloom filters have found widespread application for membership queries in areas such as networking, databases, web caching, and architectural predictions\cite{breternitz2008segmented}.
\revision{A recent model, Bloom WiSARD}\cite{Arajo2019MemoryEW}, demonstrated that replacing the RAM nodes in WiSARD with Bloom filters \revision{improves memory efficiency and model robustness} \cite{santiago2020weightless}.

Internally, a Bloom Filter is composed of $k$ distinct hash functions, each of which takes an $n$-bit input and produces an $m$-bit output, and a $2^{m}$-bit RAM.
When a new value is added to the set represented by the filter, it is passed through all $k$ hash functions, and the corresponding bit positions in the RAM are set. When the filter is checked to see if a value is in the set, the value is hashed, and the filter reports the value as present only if all $k$ of the corresponding bit positions are set.


\subsection{Bleaching}
Traditional RAM nodes activate when presented with any pattern they saw during training, even if that pattern was only seen once. This can result in overfitting, particularly for large datasets, a phenomenon known as \textit{saturation}. Bleaching \cite{bleaching} is a technique which prevents saturation by choosing a threshold $b$ such that nodes only respond to patterns they saw at least $b$ times during training. During training, this requires replacing the single-bit values in the RAM nodes with counters which track how many times a pattern was encountered. After training is complete, a bleaching threshold $b$ can be selected to maximize the accuracy of the network\footnote{Alternatively, $b$ may be chosen dynamically to serve as a tiebreaker when two or more discriminators produce an equal response. We do not explore this method of bleaching in this paper.}. 
Once $b$ has been selected, counter values greater than or equal to $b$ can be statically replaced with 1, and counter values less than $b$ with 0. Therefore, while additional memory is required during training, inference with a bleached WNN introduces no additional overhead.

In practice, bleaching can substantially improve the accuracy of WNNs. There have been several strategies proposed for finding the optimal bleaching threshold $b$; we use a binary search strategy based on the method proposed in \cite{bleaching}. Our approach performs a search between 1 and the largest counter value seen in any RAM node\revision{. Thus, both the space and time overheads of bleaching are worst-case logarithmic in the size of the training dataset.} 

\subsection{Thermometer Encoding}
Traditionally, WNNs represent their inputs as 1-bit values, where an input is 1 if it rises above some pre-determined threshold\footnote{Frequently the mean value of the input in the training data} and 0 otherwise. However, it is frequently advantageous to use more sophisticated encodings, where each parameter is represented using multiple bits\cite{wisard_encoding}. Integer encodings are not a good choice for WiSARD, since individual bits carry dramatically different amounts of information. In an 8-bit encoding, the most significant bit would carry a great deal of information about the value of a parameter, while the least significant bit would essentially be noise. Since the assignment of bits to RAM nodes is randomized, this would result in some inputs to some RAM nodes being useless.

\begin{figure}[hbtp]
\centerline{\includegraphics[width = 0.55\columnwidth]{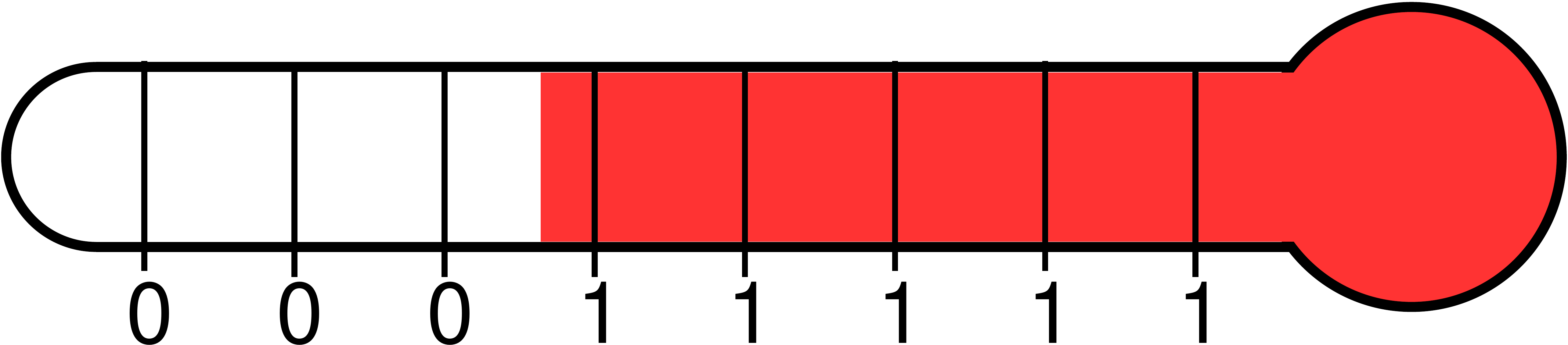}}
\caption{Like the mercury passing the gradations in a thermometer, in a thermometer encoding, bits are set to 1 from least to most significant as the encoded value increases.}
\label{fig:thermometer}
\end{figure}

In a thermometer encoding, a value is compared against a series of increasing thresholds, with the $i$'th bit of the encoded value representing the result of the comparison against the $i$'th threshold. Clearly if a value is greater than the $i$'th threshold, it is also greater than thresholds $\{0 \ldots (i-1)\}$; as Figure \ref{fig:thermometer} shows, the encoding resembles mercury passing the markings on an analog thermometer, with bits becoming set from least to most significant as the value increases. 

\section{Proposed Design: \wname}
\label{sec:proposal}

\begin{figure*}[t]
\centerline{\includegraphics[width = 0.9\textwidth]{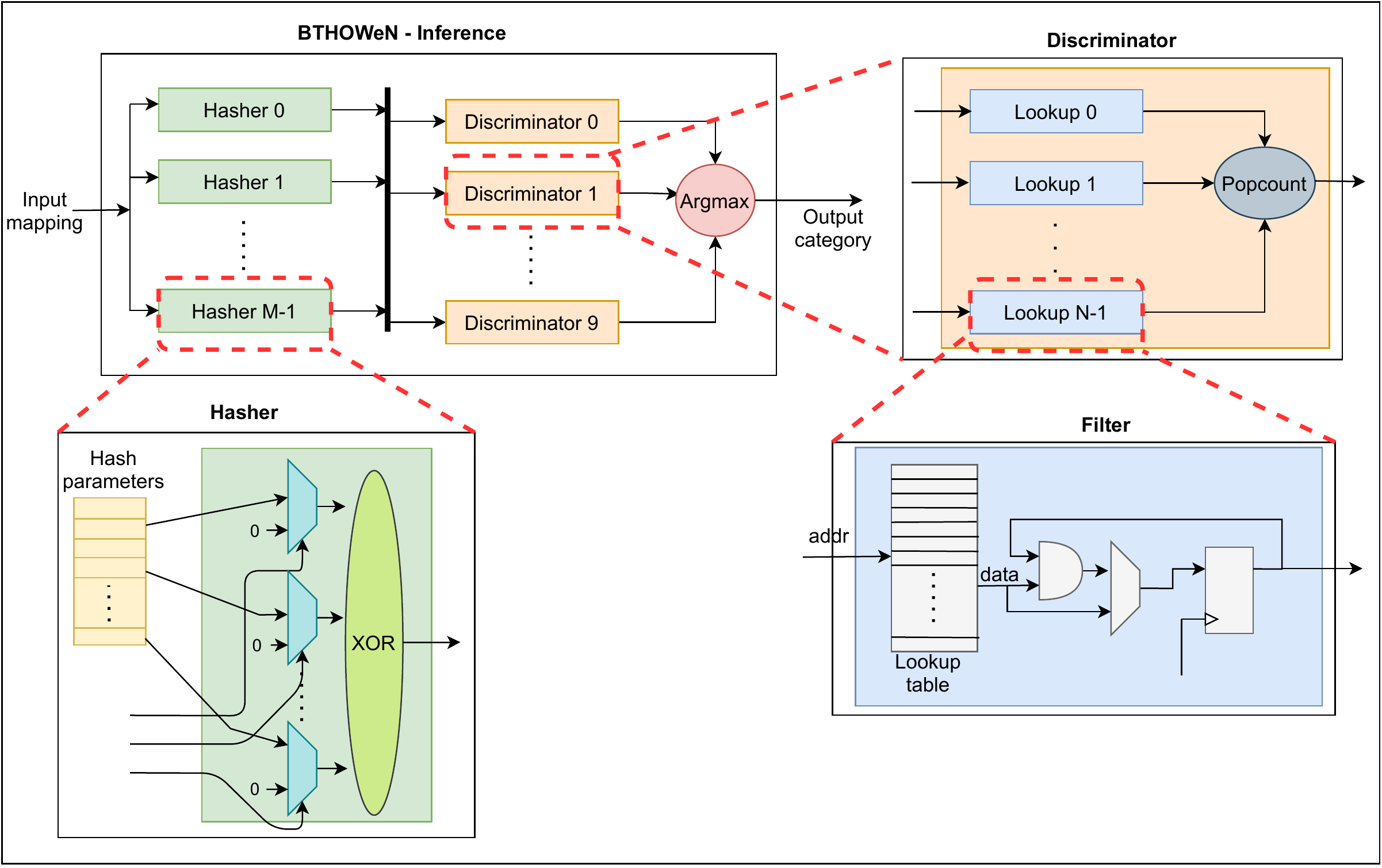}}
\caption{A diagram of the \wname~inference accelerator architecture. We divide Bloom filters into dedicated Hasher and Lookup blocks. The Hasher block computes the H3 hash function on the input data, using a shared set of random hash parameters. The Discriminator block takes hashed data as input, passes it through Lookup units, and performs a popcount on the result, returning a response. The Lookup block contains a LUT, which is accessed using the addresses produced by the hashers, and performs an \texttt{AND} reduction on the results of multiple accesses.}
\label{fig:proposed_design}
\end{figure*}

In this paper, we present \wname, a WNN architecture \revision{which improves on the prior work by incorporating} (i) counting Bloom filters to reduce model size while enabling bleaching, (ii) an inexpensive hash function which does not require arithmetic operations, and (iii) a Gaussian-based non-linear thermometer encoding to improve model accuracy.
We also present an FPGA-based accelerator for this architecture, targeting low-power edge devices, shown in Figure \ref{fig:proposed_design}.
We incorporate both hardware and software improvements over the prior work.

\subsection{Model}
\revision{Our objective is to create a hardware-aware, high-accuracy, high-throughput WNN architecture. To accomplish this goal, we enhance the techniques described in Section \ref{sec:background} with novel algorithmic and architectural improvements.}



\subsubsection{Counting Bloom Filters}
\revision{While Bloom filters were used in prior work~\cite{Arajo2019MemoryEW}, we augment them to be \textit{counting} Bloom filters.}
Bloom filters can only track whether a pattern has been seen; in order to implement bleaching, we need to know \textit{how many times} each pattern has been encountered.

\revision{A counting Bloom filter is a variant of the Bloom filter which replaces single-bit filter entries with multi-bit counters.} When an item is added to the filter, it is fed to all $k$ hash functions, and the corresponding counters are incremented. The \revision{classical} counting Bloom filter allows for items to be added to the array multiple times, and also allows for items to be removed by decrementing counters (although this introduces a risk of false negatives)\revision{\cite{electronics8070779}}.
Since we \revision{do not need} element deletion \revision{for bleaching}, we can modify the counting Bloom filter to eliminate some potential false positives. Rather than incrementing all $k$ counter values, we find the minimum of the accessed counter values, and increment only the counters which have that value. Note that \revision{false negatives are still impossible}; if a pattern has been seen $i$ times, then the smallest of its corresponding counter values must be at least $i$.

\begin{figure}[hbtp]
\centerline{\includegraphics[width = 0.9\columnwidth]{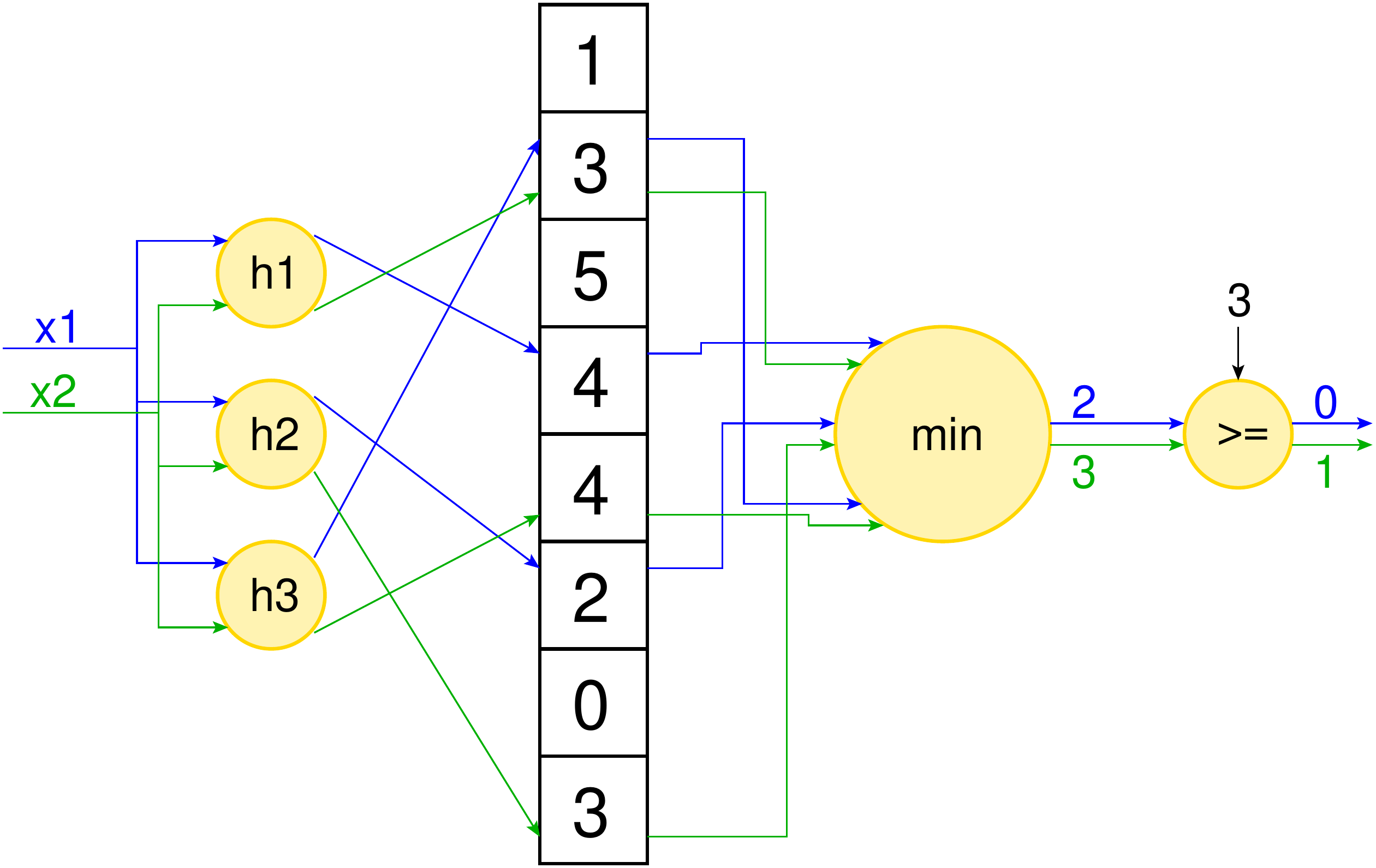}}
\caption{An example of a counting Bloom filter with $k=3$ hash functions. Hashed input $x_{1}$ corresponds to locations containing $\{3,4,2\}$; hashed input $x_{2}$ corresponds to locations containing $\{3,4,3\}$. \revision{$b = 3$} in this example (shown by the input to the ``$>=$" block), so $x_{1}$ produces an output of 0 while $x_{2}$ produces an output of 1.}
\label{fig:counting_bloom_filter}
\end{figure}

As shown in Figure \ref{fig:counting_bloom_filter}, when performing a lookup, a counting Bloom filter returns 1 if the \textit{smallest} counter value accessed is at least \revision{some threshold $b$}; thus, the possible responses become ``possibly seen at least \revision{$b$} times" and ``definitely not seen \revision{$b$} times".

\subsubsection{Hash Function Selection}
Bloom filters require multiple distinct hash functions, but do not prescribe what those hash functions should be. \revision{Prior work, including Bloom WiSARD~\cite{Arajo2019MemoryEW}\cite{santiago2020weightless}, used a double-hashing technique based on the MurmurHash\cite{MurmurHash} algorithm. However, this approach requires many arithmetic operations (e.g. 5 multiplications to hash a 32-bit value), and is therefore impractical in hardware. We identified an alternative approach based on sampling universal families of hash functions which is much less expensive to implement. Thus, while prior work used software-implemented Bloom filters, our design incorporates realistic filters which abide by hardware constraints.}

\revision{A \textit{universal family} of hash functions is a set of functions such that the odds of a hash collision are low in expectation for all functions in the family~\cite{CARTER1979143}. Some universal families consist of highly similar functions, which differ only by the choices of constant "seed" parameters.}
We considered two such families when designing \wname.

The Multiply-Shift hash family\cite{DIETZFELBINGER199719} is a universal family of non-modulo hash functions which, for an $n$-bit input size and an $m$-bit output size, implement the function \mbox{$h(x) = (ax+b) \gg (n-m)$}, where $a$ is an \textit{odd} $n$-bit integer, and $b$ is an $(n-m)$-bit integer. The Multiply-Shift hash function consists of only a few machine instructions, so is easily implemented on a CPU. \revision{However, multiplication is a relatively expensive operation in FPGAs, especially when many computations must be performed in parallel.}

\revision{By contrast, the} H3 family of hash functions\cite{CARTER1979143} requires no arithmetic operations. For an $n$-bit input \revision{$x$} and $m$-bit output, hash functions in the H3 family take the form:
\[
h(x) = x[0]p_{0} \oplus x[1]p_{1} \oplus \ldots \oplus x[n-1]p_{n-1}
\]
Here, $x[i]$ is the $i$'th bit of $x$, and $P=\{p_{0} \ldots p_{n-1}\}$ consists of $n$ random $m$-bit values. \revision{The drawback of the H3 family is that its functions} require substantially more storage for parameters when compared to the Multiply-Shift family: $nm$ bits versus just $2n-m$. 

In practice, using Bloom filters in a WiSARD model requires many independent filters, each replacing a single RAM node. Each filter in turn requires multiple hash functions. We draw all hash functions from the same universal family, and use $\mathcal{P}=\{P_{0}...P_{k-1}\}$ to represent the random parameters for a filter's $k$ hash functions.

For an implementation which uses Multiply-Shift hash functions, many multiplications need to be computed in parallel. \revision{This requires a large number of DSP slices on an FPGA}. On the other hand, when using H3 hash functions, a large register file is needed for each set of hash parameters $\mathcal{P}$. However, we observed that sharing $\mathcal{P}$ between Bloom filters did not cause any degradation in accuracy. This effectively eliminates the only \revision{comparative} disadvantage of the H3 hash function\revision{; hence,} \wname~ uses the H3 hash function with the same $\mathcal{P}$ shared between all filters.

Cryptographically-secure hash functions such as SHA and MD5 are a poor choice for Bloom filters, as their security features introduce substantial computational overhead.

\subsubsection{Implementing Thermometer Encoding}

Another enhancement we introduce in \wname~is \revision{Gaussian} non-linear thermometer encoding.
Most prior work using thermometer encodings uses equal intervals between the thresholds. The disadvantage of this approach is that a large number of bits may be dedicated to encoding outlying values, leaving fewer bits to represent small differences in the range of common values.

For thermometer encoding in \wname, we assume that each input follows a normal distribution, and compute its mean and standard deviation from training data.For a $t$-bit encoding, we divide the Gaussian into $t+1$ regions of equal probability. The values of the divisions between these regions become the thresholds we use for encoding. \revision{This provides increased resolution for values near the center of their range.}

\subsection{Training \wname}

\begin{figure}[hbtp]
\vspace{-10pt}
\centerline{\includegraphics[width = 0.6\columnwidth]{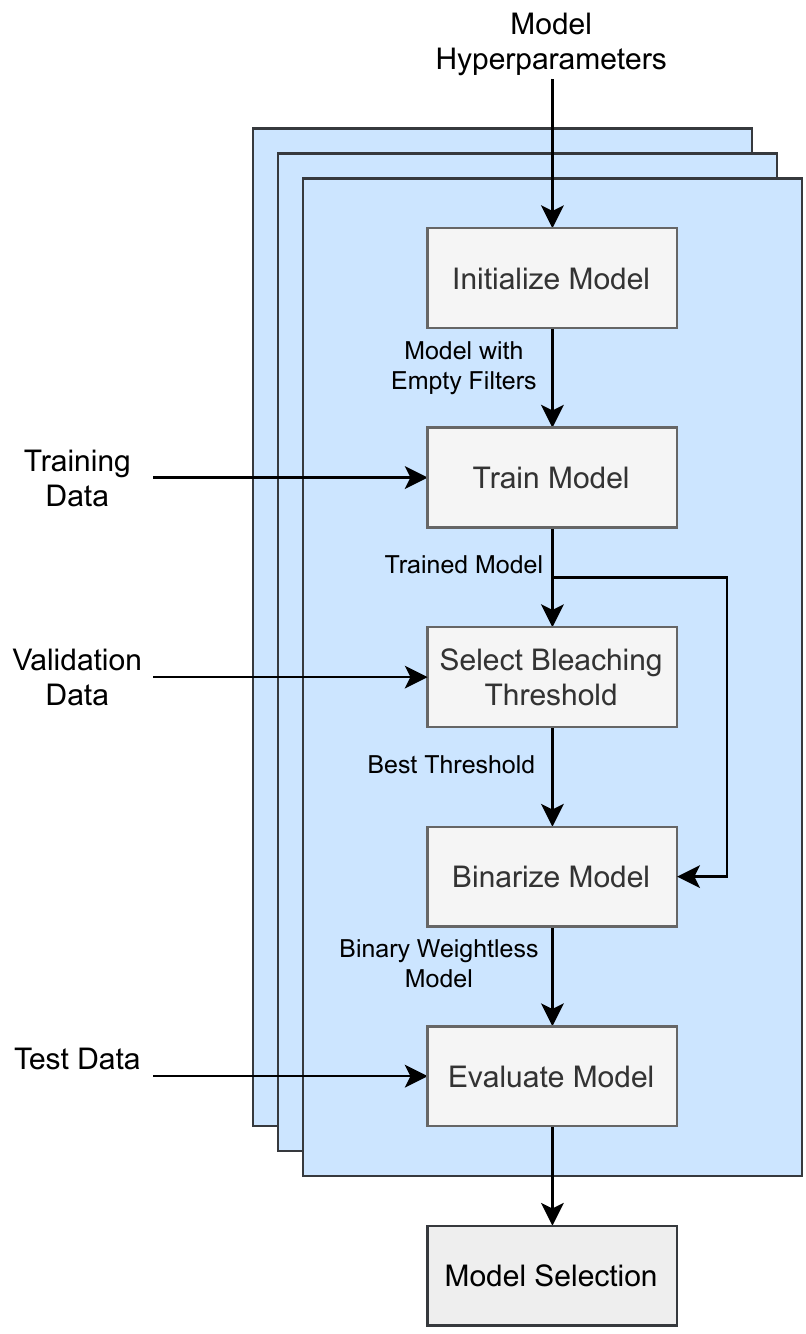}}
\caption{The training process for \wname~models. Hyperparameters, consisting of the numbers of inputs and categories for the dataset, as well as tunable parameters, are used to construct an ``empty" model, where all counter values are 0. Encoded training samples are sequentially presented to the model to update counter values. A validation set is used to select the optimal bleaching threshold $b$. This threshold is then used to binarize the trained model, replacing counters with binary values. We compare multiple models with different hyperparameters to find targets for implementation.}
\label{fig:training}
\end{figure}

The process of training a network with the \wname~ architecture is shown in Figure \ref{fig:training}. 
Hyperparameters, including the number of inputs in each sample, the number of output classes, details of the thermometer encoding, and configuration information for the Bloom filters, are used to initialize the model.

During training, samples are presented sequentially to the model. The label of the sample is used to determine which discriminator to train. The input is encoded, passed through the pseudo-random mapping, and presented to the filters in the correct discriminator. Filters hash their inputs and update their corresponding entries.

After training, the model is evaluated using the validation set at different bleaching thresholds. A binary search strategy is used to select the bleaching threshold $b$ which gives the highest accuracy. The model is then binarized by replacing filter entries less than $b$ with 0, and all other entries with 1. Binarization does not impact model accuracy, and allows counting Bloom filters to be replaced with conventional Bloom filters, which require less memory and are simpler to implement in hardware.

\vspace{-1.5mm}

\subsection{Inference with \wname}

Figure \ref{fig:proposed_design} shows the design of an accelerator for inference with \wname~ WNNs. Since reusing the same random hash parameters for all Bloom filters does not degrade accuracy, we use a central register file to hold the hash parameters. Since all discriminators receive the same inputs, Bloom filters which are at the same index but in different discriminators (e.g. filter 1 in $d_{1}$ and filter 1 in $d_{2}$) also receive identical inputs. This means that their hashed values are also identical. It is redundant and inefficient to compute the same hashed values in each discriminator. Instead, we divide the Bloom filters into separate hashing units and lookup units, where hashing units perform H3 hash operations, and lookup filters hold the Bloom filter data and perform \texttt{AND} reductions to determine the filter response. We place the hashing units at the top level of the design, before the discriminators, and broadcast their outputs to all discriminators. Since Bloom filters at the same index across different discriminators have different contents in their RAMs, lookup units can not also be shared across discriminators.

If the \revision{bus bringing data from off-chip has insufficient bandwidth}, then the accelerator will finish before the next input is ready.
In this case, we can reduce the number of hash units by having each one compute the hashed inputs for multiple lookup units. We store partial results in a large central register until all hashes have been computed for a single set of hash parameters, then pass the hash results to all filters simultaneously, ensuring they operate in lockstep. This strategy allows us to reduce the area of the design without decreasing effective throughput.

The popcount module counts the number of 1s in the outputs of the filters in a discriminator, and the argmax module determines the index of the discriminator with the strongest response. These are equivalent to the corresponding modules in a conventional WiSARD model.

Since training with bleaching requires multi-bit counters for each entry in each Bloom filter, it introduces a large amount of memory overhead. For instance, in our experimentation, we found that some models had optimal bleaching values of more than 400. If we used saturating counters large enough to represent this value in the accelerator, it would increase the memory usage of the design by a factor of 9. Since our accelerator is intended for use in low-power edge devices, the advantages of supporting on-chip training do not seem worth the cost.



\section{Evaluation Methodology}
\label{sec:methodology}



\subsection{Hardware Implementation}
Our hardware source is written using Mako-templated SystemVerilog. Mako is a template library for Python which allows the Python interpreter to be run as a preprocessing step for arbitrary languages. This allows for greater flexibility and ease of use than the SystemVerilog preprocessor alone. When Mako is invoked, it generates pure SystemVerilog according to user-specified design parameters.

We targeted two different Xilinx FPGAs for this project. For most designs, we used the xc7z020clg400-1 FPGA, a small, inexpensive FPGA available in the Zybo Z7 development board, which was used for prior work \cite{wisard_fpga}. For our largest design, we targeted the Kintex UltraScale xcku035-ffva1156-1-c. Timing, power, and area numbers were obtained from Xilinx Vivado. 
The choice by the prior work to profile the entire system revealed a major bottleneck in the form of SD card read bandwidth \cite{wisard_fpga}; we are interested in the performance of the accelerator itself.
We implement all models with a 100 MHz clock rate, and collect power numbers assuming a 12.5\% switching rate. 

Hash units produce output at a maximum throughput of 1 hash/cycle. Lookup units can consume hashed inputs at a rate of 1/cycle, and produce output at a rate of 1/$k$ cycles, where $k$ is the number of hash functions associated with a Bloom filter. Therefore, there is no point in having more hashing units than lookup units, and the maximum throughput of the design is 1/$k$ cycles. This throughput could be improved by allowing multiple addresses to be read simultaneously in the lookup units. However, this would greatly increase circuit area, and such a design would likely not be any faster in practice; $k$ is typically small enough that reading data into the accelerator will almost always be the bottleneck.


At the top level of the design, we use a double-buffered deserialization unit which accumulates input data from the bus until a full sample has been read, then passes the entire sample to the accelerator. This helps enable all hardware units to operate in lockstep, simplifying our state machine logic and verification effort.


\subsection{Datasets and Training}
We created models for all classifier datasets discussed in \cite{Arajo2019MemoryEW}: MNIST \cite{lecun-mnisthandwrittendigit-2010}, Ecoli \cite{ecoli_dataset}, Iris\cite{iris_dataset},  Letter\cite{letter_dataset}, Satimage\cite{satimage_dataset}, Shuttle\cite{shuttle_dataset},  Vehicle\cite{vehicle_dataset}, Vowel\cite{vowel_dataset}, and Wine\cite{wine_dataset}.
Since our accelerator does not support on-chip training, we implemented the training of models in software. This was done in Python, using the Numba JIT compiler to reduce the runtime of performance-critical functions. We performed a 90-10 train/validation split on the input dataset, using the former to learn the values in the counting Bloom filters and the latter to set the bleaching threshold.

\subsection{WNN Model Sweeping}

There are several model hyperparameters which can be changed to impact the size and accuracy of the model. Increasing the size of the Bloom filters decreases the likelihood of false positives, and thus improves accuracy. However, this greatly increases the model size, and eventually provides diminishing returns for accuracy as false positives become too rare to matter. Increasing the number of input bits to each Bloom filter broadens the space of Boolean functions the filters can learn to approximate, and makes the model size smaller as fewer Bloom filters are needed in total. However, it also increases the likelihood of false positives, since more unique patterns are seen by each filter. Increasing the number of hash functions per Bloom filter can improve accuracy up to a point, but past a certain point actually begins to \textit{increase} the frequency of false positives \cite{BloomFilter}. Lastly, increasing the number of bits in the thermometer encoding can improve accuracy at the cost of model size, but again provides diminishing returns as the amount of information each bit conveys decreases.

In order to identify optimal model hyperparameters, we ran many different configurations in parallel using an automated sweeping methodology. For MNIST, we used 1008 distinct configurations, sweeping all combinations of the hyperparameter settings shown in Table \ref{tab:sweep_params}. For smaller datasets, we explored using 1-16 encoding bits per input, 128-8192 entries per Bloom filter, 1-6 hash functions per filter, and 6-64 input bits per Bloom filter.

\begin{table}[h]
\centering
\caption{Hyperparameters swept for the creation of \wname~models for the MNIST dataset}
\begin{adjustbox}{center}
\begin{tabular}{|c|c|} 
 \hline
 \rowcolor{LightBlue}
 Hyperparameter & Values \\
 \hline
 Encoding Bits per Input & 1, 2, 3, 4, 5, 6, 7, 8\\ 
 \hline
 Input Bits per Bloom Filter & 28, 49, 56\\
 \hline
 Entries per Bloom Filter & 128, 256, 512, 1024, 2048, 4096, 8192\\
 \hline
 Hash Functions per Bloom Filter & 1, 2, 3, 4, 5, 6\\
 \hline
\end{tabular}
\end{adjustbox}
\label{tab:sweep_params}
\end{table}

\vspace{-1em}
\subsection{DNN Model Identification and Implementation}
For each dataset, we trained MLPs that had similar accuracy to our \wname~ models. \revision{We identified the smallest iso-accuracy MLPs using a hyperparameter sweep.}
The trained models were then quantized to 8-bit precision to generate a TensorFlow Lite model. Hardware was generated for each \revision{MLP} using the \texttt{hls4ml} tool \cite{hls4ml}. \texttt{hls4ml} takes 4 inputs: (1) the weights generated by TensorFlow (.h5 format), (2) the structure of the model generated by TensorFlow (.json format), (3) the precision to be used for the hardware, and (4) the FPGA part being targeted. It generates C++ code corresponding to the model, and then invokes Xilinx Vivado HLS to generate the hardware design. We modified the generated C++ code such that the I/O interface width matched that of our hardware design for WNNs in order to ensure a fair comparison.
We also modified HLS pragmas as needed to ensure that the resultant RTL could fit on the Zybo FPGA.
The hardware design generated by Vivado HLS (invoked by \texttt{hls4ml}) was then synthesized and implemented using Xilinx Vivado to obtain area, latency, and power consumption metrics.

For the MNIST dataset, in addition to MLPs, we compared the \wname~implementation with comparably accurate CNNs based on the LeNet-1 \cite{lecun-mnisthandwrittendigit-2010} architecture. The LeNet-1 implementations generated by \texttt{hls4ml} consume an order of magnitude more area and energy than optimized implementations reported in literature \cite{lenet_singapore}. Therefore, \revision{in order to make a more fair comparison}, we used the latency and resource numbers for optimized implementations reported by Arish et. al. \cite{lenet_singapore}. We then used the Xilinx Power Estimator (XPE) \cite{xpe} to get approximate power values \revision{for the CNN.}

\section{Results}
\label{sec:results}

\subsection{Selected \wname~Models}
After performing a hyperparameter sweep, we needed to select one or more trained models for FPGA implementation, balancing tradeoffs between model size and accuracy. For each dataset except for MNIST, there was one model which was very clearly the best, with all more accurate models being many times larger.

Since MNIST is a more complex dataset, there was no clear single    ``best" model - instead, we identified ``Small", ``Medium", and ``Large" models, which balanced size and accuracy at different points. Our objectives for the three MNIST models were:
\begin{itemize}
    \item The small model would be comparable in area to the prior \revision{FPGA model in \cite{wisard_fpga}}
    \item The medium would be larger, but could still fit on the same FPGA (i.e. the Zybo Z7 board)
    \item The large model would fit on a mid-size commercial FPGA
\end{itemize}
\revision{We also experimented with MNIST models using traditional linear thermometer encodings, and observed a 12.9\% reduction in mean error using the Gaussian encoding.}

The configurations for all the models we selected are shown in Table \ref{tab:model_specs}.



\begin{table}[htbp]
\centering
\caption{Details of the selected \wname~ models}
\begin{adjustbox}{center}
\begin{tabular}{|c|c|c|c|c|c|c|} 
 \hline
 \rowcolor{LightBlue} \centering
 Model & Bits & Bits & Entries &  Hashes & Size & Test \\
 \rowcolor{LightBlue} \centering
 Name & /Input & /Filter & /Filter & /Filter & (KiB) & Acc.\\
 \hline
 MNIST-Small & 2 & 28 & 1024 & 2 & 70.0 & 0.934\\
 \hline
 MNIST-Medium & 3 & 28 & 2048 & 2 & 210 & 0.943\\
 \hline
 MNIST-Large & 6 & 49 & 8192 & 4 & 960 & 0.952\\
 \hline
 Ecoli & 10 & 10 & 128 & 2 & 0.875 & 0.875\\
 \hline
 Iris & 3 & 2 & 128 & 1 & 0.281 & 0.980\\
 \hline
 Letter & 15 & 20 & 2048 & 4 & 78.0 & 0.900\\
 \hline
 Satimage & 8 & 12 & 512 & 4 & 9.00 & 0.880\\
 \hline
 Shuttle & 9 & 27 & 1024 & 2 & 2.63 & 0.999\\
 \hline
 Vehicle & 16 & 16 & 256 & 3 & 2.25 & 0.762\\
 \hline
 Vowel & 15 & 15 & 256 & 4 & 3.44 & 0.900\\
 \hline
 Wine & 9 & 13 & 128 & 3 & 0.422 & 0.983\\
 \hline
\end{tabular}
\end{adjustbox}
\label{tab:model_specs}
\end{table}

\vspace{-1em}
\subsection{Comparison with \revision{Iso-Accuracy~}Deep Neural Networks}

\begin{table*}[htb]
\centering
\caption{Comparison of \wname~FPGA Models with Quantized DNNs of similar accuracy implemented in FPGAs. CNNs for MNIST are LeNet-1 variations from ~\cite{lenet_singapore}. WNN and MLP for each data set is grouped in nearby rows for easy comparison.}
\begin{adjustbox}{center}
\begin{tabular}{|c|c|p{0.8cm}|p{0.7cm}|p{0.7cm}|p{1.6cm}|p{1.7cm}|c|c|p{0.8cm}|c|c|} 
 \hline
 \rowcolor{LightBlue} \centering
 \textbf{Dataset} & \textbf{Model} & \textbf{Bus Width} & \textbf{Cycles per Inf.} & \textbf{Hash Units} & \textbf{Dyn. Power (Tot. Power) (W)} & \textbf{Dyn. Energy (Tot. Energy) (nJ/Inf.)} & \textbf{LUTs} & \textbf{FFs} & \textbf{BRAMs (36Kb)} & \textbf{DSPs} & \textbf{Accuracy} \\
 \hline
 \multirow{8}{*}{MNIST} & \wname-Small & 64 & 25 & 5 & 0.195 (0.303) & 48.75 (75.8) & 15756 & 3522 & 0 & 0 & 0.934 \\
 \cline{2-12}
   & \wname-Medium & 64 & 37 & 5 & 0.386 (0.497) & 142.8 (183.9) & 38912 & 6577 & 0 & 0 & 0.943\\
 \cline{2-12}
   & \wname-Large & 64 & 74 & 6 & 3.007 (3.509) & 2225 (2597) & 151704 & 18796 & 0 & 0 & 0.952\\
 \cline{2-12}
   & \wname-Large* & 256 & 19 & 24 & 3.158 (3.661) & 600.0 (695.6) & 158367 & 25905 & 0 & 0 & 0.952\\
 \cline{2-12}

 \cline{2-12}
  & MLP 784-16-10 & 64 & 846 & - & 0.029 (0.134) & 245 (1133) & 2163 & 3007 & 8 & 28 & 0.946\\
 \cline{2-12}
  & CNN 1 (LeNet1)\cite{lenet_singapore} & 64 & 33615 & - & 0.058 (0.163) & 19497 (54792) & 5753 & 3115 & 7 & 18 & 0.947\\
 \cline{2-12}
  & CNN 2 (LeNet1)\cite{lenet_singapore} & 64 & 33555 & - & 0.043 (0.148) & 14429 (49661) & 3718 & 2208 & 5 & 10 & 0.920\\
 \cline{2-12}
 
 \cline{2-12}
   & Hashed WNN \cite{wisard_fpga} & 32 & 28 & - & 0.423 (0.528) & 118.4 (147.8) & 9636 & 4568 & 128.5 & 5 &  0.907   \\
  
 \hline
 \rowcolor{light-gray} 
 \hline
  & \wname~& 64 & 2 & 7 & 0.012 (0.117) & 0.24 (2.34) & 353 & 223 & 0 & 0 & 0.875 \\
 \cline{2-12}
 \rowcolor{light-gray} 
 \multirow{-2}*{Ecoli} & MLP 7-8-8 & 64 & 14 & - & 0.03 (0.135) & 4.2 (18.9) & 1596 & 1615 & 0 & 0 & 0.875\\
 \hline
  
 \hline
 \multirow{2}*{Iris} & \wname~& 64 & 1 & 6 & 0.005 (0.109) & 0.05 (1.09) & 57 & 90 & 0 & 0 & 0.980\\
 \cline{2-12}
  & MLP 4-4-3 & 64 & 10 & - & 0.008 (0.112) & 0.8 (11.2) & 427 & 488 & 0 & 0 & 0.980\\
 \hline
  
 \hline
 \rowcolor{light-gray} 
  & \wname~& 64 & 4 & 12 & 0.623 (0.738) & 24.92 (29.52) & 21603 & 2715 & 0 & 0 & 0.900\\
 \cline{2-12}
 \rowcolor{light-gray} 
 \multirow{-2}*{Letter} & MLP 16-40-26 & 64 & 26 & - & 0.109 (0.259) & 39.52 (67.34) & 17305 & 15738 & 0 & 0 & 0.904\\
 \hline

 \hline
 \multirow{2}*{Satimage} & \wname~& 64 & 5 & 24 & 0.084 (0.190) & 4.2 (9.5) & 3771 & 1131 & 0 & 0 & 0.880\\
 \cline{2-12}
  & MLP 36-16-16-6 & 64 & 25 & - & 0.039 (0.144) & 9.75 (36) & 7007 & 7558 & 0 & 0 & 0.878\\
 \hline

 \hline 
 \rowcolor{light-gray} 
  & \wname~& 64 & 2 & 3 & 0.018 (0.123) & 0.36 (2.46) & 593 & 121 & 0 & 0 & 0.999\\
 \cline{2-12}
 \rowcolor{light-gray} 
 \multirow{-2}*{Shuttle} & MLP 9-4-7 & 64 & 14 & - & 0.013(0.118) & 1.82 (16.52)  & 693 & 711  & 0 & 0 & 0.999\\
 \hline

 \hline
 \multirow{2}*{Vehicle} & \wname~& 64 & 5 & 18 & 0.038 (0.143) & 1.9 (7.15) & 1781 & 597 & 0 & 0 & 0.762\\
 \cline{2-12}
  & MLP 18-16-4 & 64 & 15 & - & 0.024 (0.128) & 3.6 (19.2) & 2824  & 3035 & 0 & 0 & 0.766\\
 \hline

 \hline
 \rowcolor{light-gray} 
  & \wname~& 64 & 2 & 12 & 0.040 (0.145) & 0.8 (2.9) & 1559 & 756 & 0 & 0 & 0.900\\
 \cline{2-12}
 \rowcolor{light-gray} 
 \multirow{-2}*{Vowel} & MLP 10-18-11 & 64 & 18 & - & 0.07 (0.175) & 12.6 (31.5) & 5743 & 4663 & 0 & 0 & 0.903\\
 \hline

 \hline
 \multirow{2}*{Wine} & \wname~& 64 & 3 & 9 & 0.012 (0.117) & 0.36 (3.51) & 585 & 239 & 0 & 0 & 0.983\\
 \cline{2-12}
  & MLP 13-10-3 & 64 & 14 & - & 0.026 (0.131) & 3.64 (18.34) & 1836 & 1832 & 0 & 0 & 0.983\\
 \hline
\end{tabular}
\end{adjustbox}
\label{tab:model_results}
\vspace{-10pt}
\end{table*}

Table \ref{tab:model_results} shows FPGA implementation results for \wname~ models and \revision{iso-accuracy} quantized DNNs \revision{identified using a hyperparameter sweep}~across the nine datasets.
For the MNIST dataset, the medium \wname~ model is only 0.3\% less accurate than the MLP, consumes just 16\% of the energy of the MLP model, and reduces latency by almost 96\%. The MLP uses fewer LUTs and FFs than the medium \wname~ model, but also requires DSP blocks and BRAMs on the FPGA. The \wname~ model compares even more favorably against CNNs. For example, CNN-1 has an accuracy of 94.7\%, which is only slightly better than the 94.3\% accuracy of the medium \wname~ model. But even with a pipelined CNN implementation, \wname~ consumes less than 0.4\% of the energy of the CNN, while reducing latency from 33.6k cycles to just 37.

As Table \ref{tab:model_results} illustrates, for all datasets except MNIST and Letter, the \wname~ model's hardware implementation consumes less resources (LUTs and FFs) than its MLP counterpart. The reduction in total energy consumption of the \wname~ models ranges from 56.2\% on Letter to 90.8\% on Vowel. Reduction in latency ranges from 66.7\% on Letter to 90.0\% on Iris.

\begin{figure*}[t]
\vspace{-3pt}
\centerline{\includegraphics[width =\textwidth]{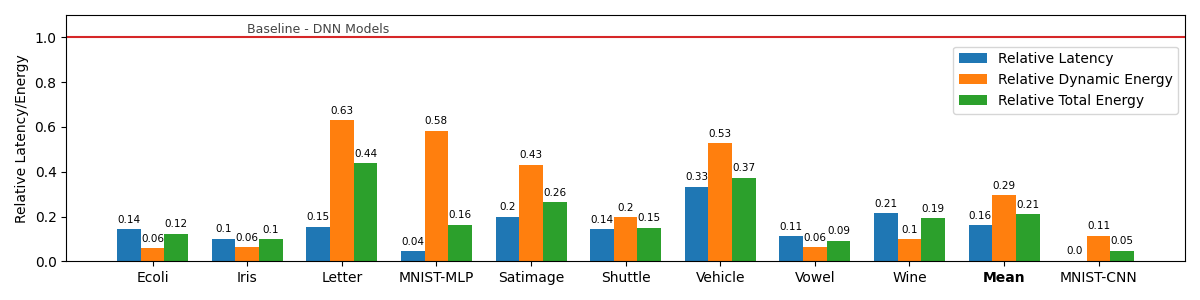}}
\vspace{-7pt}
\caption{\revision{The relative latencies and energies of} \wname~models  \revision{versus iso-accuracy DNNs}. For MNIST, our Medium model is compared against the MNIST MLP model, and our Large (64b bus) model is compared against CNN 1. The results of the comparison with CNN 1 are not used in the computation of the average, since the Large model uses a different FPGA. \revision{We implemented baseline MLPs and~\wname models at 100MHz and obtained metrics from Xilinx tools.}}
\label{fig:dnn_comparison}
\end{figure*}

Figure \ref{fig:dnn_comparison} summarizes these results, showing the relative latencies, dynamic energies, and total energies of \wname~ models compared to DNNs.
Overall, \wname~ models are significantly faster and more energy efficient than DNNs of comparable accuracy.

The different-sized models for MNIST, shown in Table \ref{tab:model_results}, provide multiple tradeoff points for energy and accuracy.
The Small and Medium models provide good energy efficiency, though the Medium model uses over twice the energy for a 0.9\% improvement in accuracy. The Large model does not fit on the Zybo FPGA, so is implemented on a larger FPGA with much higher static power consumption, and is much slower and less energy-efficient. However, if we take advantage of the large number of I/O pins on the large FPGA and implement a 256b bus, energy consumption is reduced by nearly a factor of 4 (the ``Large*" row in Table \ref{tab:model_results}).

\subsection{Comparison with Prior Weightless Neural Networks}
\revision{Bloom WiSARD, the prior state-of-the-art for WNNs, used Bloom filters to achieve far smaller model sizes than conventional WiSARD models with only slight penalties to accuracy \cite{Arajo2019MemoryEW}. Results were reported on nine multi-class classifier datasets, which we adopted for our analysis.}

\begin{figure*}[t]
\vspace{-3pt}
\centerline{\includegraphics[width =\textwidth]{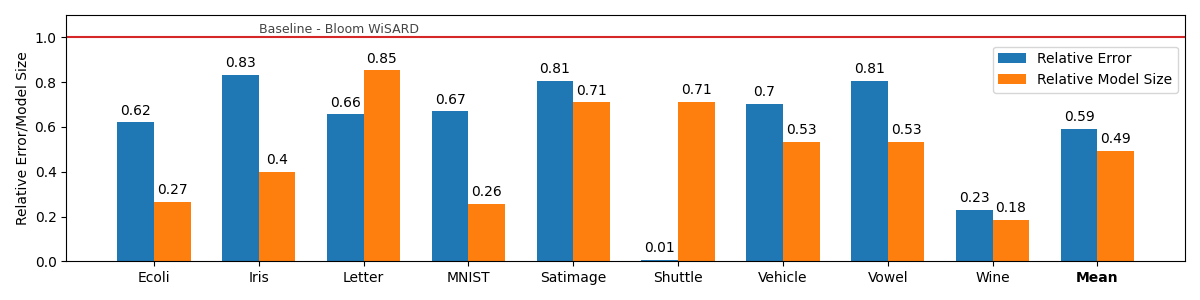}}
\vspace{-7pt}
\caption{\revision{The} relative errors and model sizes of the models shown in Table \ref{tab:model_specs} versus \revision{Bloom WiSARD} \cite{Arajo2019MemoryEW}. \wname~outperforms the prior work on all nine datasets in both accuracy and model size. For the MNIST dataset, our MNIST-Medium model was used for comparison.}
\label{fig:bloom_wisard_comparison}
\vspace{-10pt}
\end{figure*}

We \revision{compared the} \wname~models in Table \ref{tab:model_specs} against the \revision{results reported by Bloom WiSARD on} all nine datasets, achieving superior accuracy with a smaller model parameter size in all cases. Details are shown in Table \ref{tab:versus-bloom} and summarized in Figure \ref{fig:bloom_wisard_comparison}. On average, our models have 41\% less error with a 51\% smaller model size \revision{compared to Bloom WiSARD, which did not incorporate bleaching or thermometer encoding. Our improvements indicate the benefits of these techniques.} This comparison is done for software only, since the prior work did not have a hardware implementation. However, we anticipate that our advantage in hardware would be even larger due to our much simpler and more efficient choice of hash function.

One unusual result is on the Shuttle dataset, for which our model has \textasciitilde99\% less error than prior work. Shuttle is an anomaly-detection dataset in which 80\% of the training data belongs to the ``normal" class\cite{shuttle_dataset}\revision{. We suspect that, since Bloom WiSARD does not incorporate bleaching, the discriminator corresponding to this class became saturated during training.}

\begin{table}[htbp]
\centering
\caption{Accuracy and model size (in KiB) of proposed \wname~vs best memory-efficient prior work \revision{(Bloom WiSARD)~\cite{Arajo2019MemoryEW}}}
\begin{adjustbox}{center}
\begin{tabular}{|p{1.9cm}|p{1.2cm}|p{1.2cm}|p{1.1cm}|p{1.1cm}|} 
 \hline
 \rowcolor{LightBlue}
 \textbf{Model Name} & \textbf{Accuracy} & \textbf{Accuracy} & \textbf{Size} & \textbf{Size} \\
  \rowcolor{LightBlue}
   & \textbf{\revision{(Bloom}} & \textbf{(This} & \textbf{\revision{(Bloom}} & \textbf{(This} \\
    \rowcolor{LightBlue}
    & \textbf{\revision{WiSARD)}} & \textbf{work)} & \textbf{\revision{WiSARD)}} & \textbf{work)} \\
 \hline
 MNIST-Small & \multirow{3}{*}{\revision{0.915}} & 0.934 & \multirow{3}{*}{\revision{819}} & 70.0\\
 \cline{1-1} \cline{3-3} \cline{5-5}
 MNIST-Medium & & 0.943 & & 210\\
 \cline{1-1} \cline{3-3} \cline{5-5}
 MNIST-Large & & 0.952 & & 960\\
 \hline
 Ecoli & 0.799 & 0.875 & \revision{3.28} & 0.875\\
 \hline
 Iris & 0.976 & 0.980 & 0.703 & 0.281\\
 \hline
 Letter & 0.848 & 0.900 & \revision{91.3} & 78.0\\
 \hline
 Satimage & 0.851 & 0.880 & \revision{12.7} & 9.00\\
 \hline
 Shuttle & 0.868 & 0.999 & \revision{3.69} & 2.63\\
 \hline
 Vehicle & 0.662 & 0.762 & \revision{4.22} & 2.25\\
 \hline
 Vowel & 0.876 & 0.900 & \revision{6.44} & 3.44\\
 \hline
 Wine & 0.926 & 0.983 & \revision{2.28} & 0.422\\
 \hline
\end{tabular}
\end{adjustbox}
\label{tab:versus-bloom}
\vspace{-15pt}
\end{table}

\subsection{Comparison with Prior FPGA Implementation}
In \cite{wisard_fpga}, a WNN accelerator for MNIST was implemented on the Zybo FPGA \revision{(xc7z020clg400-1) with Vivado HLS. We used this same FPGA at the same frequency (100 MHz).}
The row with Model=``Hashed WNN" in Table \ref{tab:model_results}  shows the implementation results for the prior art accelerator. Its latency and energy consumption are between our small and medium models, but it is much less accurate than even our small model. This accelerator is greatly harmed by its slow memory access, which increases the impact of static power on its energy consumption.

\revision{Exact accelerator latency values were not published.} The accelerator reads in one 28-bit filter input per cycle, and uses a 1-bit-per-input encoding, so it takes 28 cycles to read in a 784-bit MNIST sample. Therefore, \revision{we use} 28 cycles as a lower bound on the time per inference for their design. The energy number in Table \ref{tab:model_results} for \cite{wisard_fpga} is a lower bound based on this cycle count \revision{and published power values}.

Our implementation \revision{has significant differences which contribute to ~\wname's superior accuracy and efficiency}:
\begin{itemize}
    \item The prior accelerator \revision{used a simple hash-table-based encoding scheme which had explicit hardware for collision detection; we use an approach based on counting Bloom filters which does not need collision detection.}
    \item Models for the prior accelerator did not incorporate bleaching or thermometer encoding; instead, they used a simple 1-bit encoding based on comparison with a parameter's mean value. \revision{We use counting Bloom filters to enable bleaching.}
\end{itemize}
\revision{Since the prior accelerator~\cite{wisard_fpga} did not incorporate bleaching, training did not require multi-bit counters, making it inexpensive to support.}

\subsection{Model Tradeoff Analysis}
\begin{figure*}[htb]
    \vspace{-7pt}
    \centering
    \begin{adjustbox}{center}
    \begin{tabular}{ccc}
    \revision{A}\includegraphics[width=0.28\paperwidth]{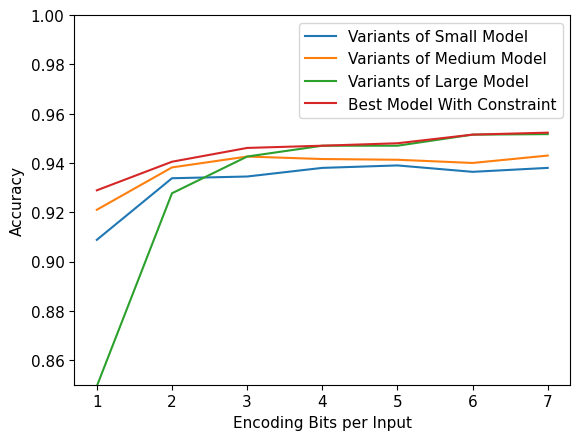} &   \revision{B}\includegraphics[width=0.28\paperwidth]{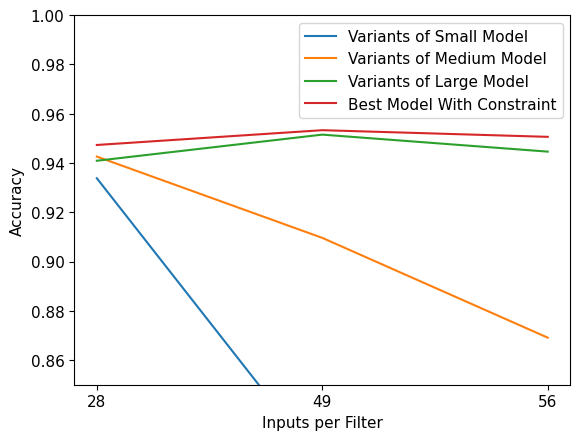} &
    \revision{C}\includegraphics[width=0.28\paperwidth]{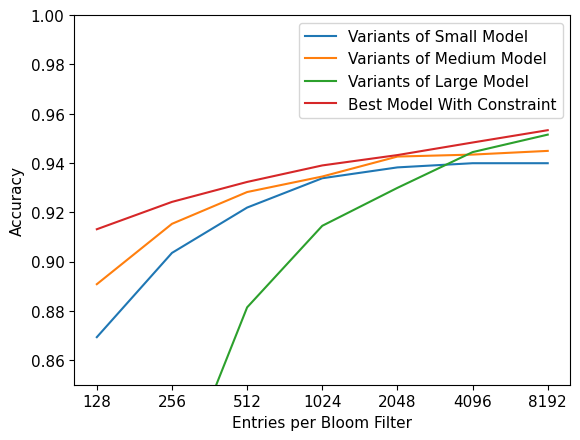} \\
    \revision{D}\includegraphics[width=0.28\paperwidth]{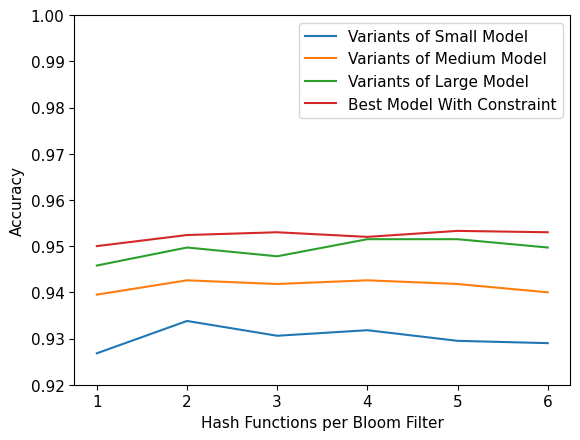} &
    \revision{E}\includegraphics[width=0.26\paperwidth]{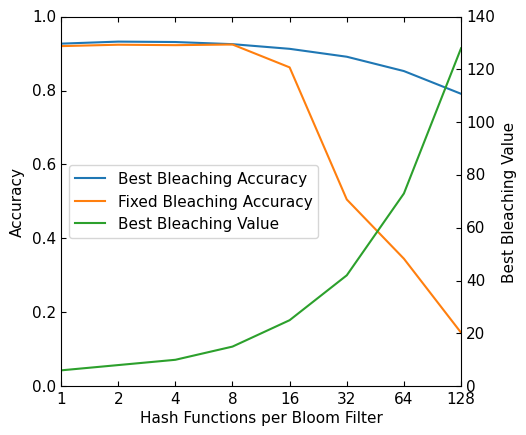} &
    \revision{F}\includegraphics[width=0.28\paperwidth]{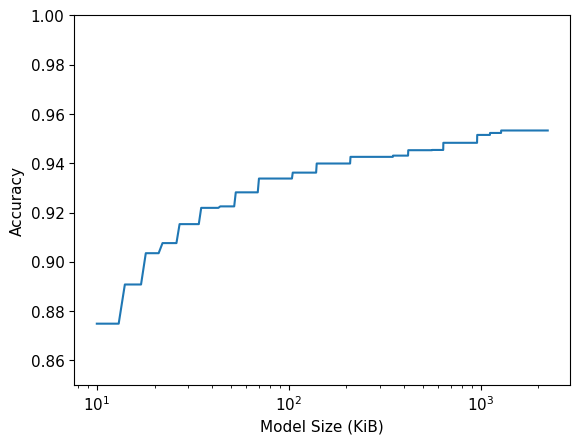} \\
    \end{tabular}
    \end{adjustbox}
    \caption{Sweeping results for MNIST with the configurations described in Table \ref{tab:sweep_params}, showing the impact of (a) the number of bits used to encode each input on accuracy, (b) the number of inputs to each Bloom filter on accuracy, (c)  the number of entries in the LUTs of each Bloom filter on accuracy, (d) the number of distinct hash functions per Bloom filter on accuracy, and (e) large numbers of hash functions on accuracy with a fixed vs. varying bleaching value. Subfigure (f) shows the most accurate model which we could obtain under a given maximum model size. }
    \label{fig:sweep_results}
    \vspace{-10pt}
\end{figure*}

We use MNIST as an illustrative example of the tradeoffs present in model selection. Figure \ref{fig:sweep_results} presents the results obtained from sweeping over the MNIST dataset with the configurations presented in Table \ref{tab:sweep_params}. 
In the first four subplots of Figure \ref{fig:sweep_results}, we vary one hyperparameter of the model, respectively, the number of bits used to encode each input, the number of inputs to each Bloom filter, the number of entries in each filter, and the number of distinct hash functions for each filter. We show four lines: three of them represent the Small, Medium, and Large models where only the specified hyperparameter was varied, while the fourth represents the best model with the given value for the hyperparameter.

We see diminishing returns as the number of encoding bits per input and the number of entries per Bloom filter increase. The Small and Medium models rapidly lose accuracy as the number of inputs per filter increases, but the Large model, with its large filter LUTs, is able to handle 49 inputs per filter without excessive false positives. These results align with the theoretical behaviors discussed earlier.

One surprising result was that, although there was a slight accuracy increase going from 1 hash function per filter to 2, continuing to increase this had minimal impact. In theory, we would expect that continuing to increase this value would eventually result in a loss of accuracy due to high false positive rates. One explanation for this is that the \wname~ model reports the index of the class with the strongest response; since a higher false positive rate would impact the response of \textit{all} classes, the predicted class should remain unaffected as long as the increase is proportional. Another observation, shown in the fifth subplot of Figure \ref{fig:sweep_results}, is that the optimal bleaching value $b$ increases to compensate for the larger number of hash functions. This plot shows variants of the Small model with up to 128 hash functions per Bloom filter. When $b$ is fixed at 16, accuracy collapses, but when the optimal value is chosen using the same binary search strategy we use normally, it is better able to compensate. This provides a good example of how bleaching improves the robustness of \wname.

The last subplot shows the most accurate MNIST model we were able to obtain with a given maximum model size. We notice diminishing returns as model size increases. It is evident that in order to exceed 96\% accuracy with reasonable model sizes, additional algorithmic improvements will be needed.


\section{Conclusion}
\label{sec:conclusion}

While most machine learning research centers around DNNs, we explore an alternate neural model, the Weightless Neural Network, for edge inference. We incorporate enhancements such as counting Bloom filters, inexpensive H3 hash functions and a Gaussian-based non-linear thermometer encoding into the WiSARD weightless neural model\revision{, improving state-of-the-art WNN MNIST accuracy~\cite{Arajo2019MemoryEW} from 91.5\% to 95.2\%}. The proposed \wname~ architecture is compared to state-of-the-art weightless models as well as MLPs and CNNs of similar accuracy. An FPGA accelerator for \wname~ is also presented. Compared to prior WNNs, \wname~ reduces error by 41\% and model size by 51\% across nine datasets. Compared to \revision{iso-accuracy} MLP models, \wname~ consumes \textasciitilde20\% of the total energy while reducing latency by \textasciitilde85\%. \revision{Energy/latency improvements over CNNs are even larger, although CNNs have higher accuracy.}

There are many opportunities for future work in this domain. 
There are algorithmic improvements we would like to explore, including weightless convolutional neural networks, better input remapping, and converting pretrained DNNs to WNNs. \revision{
Preliminary experiments suggest that backpropagation-based training approaches can significantly improve WNN model accuracy, making them feasible for broader applications.}

We believe that WNNs hold substantial promise for inference on the edge. While WNNs have historically trailed in accuracy to DNNs,  algorithmic improvements such as bleaching demonstrate that accuracy and efficiency can go up with enhanced architectures and training techniques. The low latency and low energy benefits that can be obtained from WNNs warrant further research in this area.


\bibliographystyle{IEEEtran}
\bibliography{bibliography}

\end{document}